\documentclass[11pt]{article}

\usepackage{epsfig,amssymb}
\usepackage{fullpage}
\usepackage{graphicx}
\usepackage{color}

\usepackage{subfigure}

\begin{document}

\title{Amplitude death in systems of coupled oscillators with distributed-delay coupling}

\author{Y.N. Kyrychko\thanks{Corresponding author. Email: y.kyrychko@sussex.ac.uk},\hspace{0.2cm} K.B. Blyuss
\\\\ Department of Mathematics, University of Sussex,\\Brighton, BN1 9QH, United Kingdom\\\\
\and E. Sch\"oll\\\\
Institut f\"ur Theoretische Physik, Technische Universit\"at Berlin,\\ 10623 Berlin, Germany}

\maketitle

\begin{abstract}
This paper studies the effects of coupling with distributed delay on the suppression of oscillations in a system of coupled Stuart-Landau oscillators. Conditions for amplitude death are obtained in terms of strength and phase of the coupling, as well as the mean time delay and the width of the delay distribution for uniform and gamma distributions. Analytical results are confirmed by numerical computation of the eigenvalues of the corresponding characteristic equations. These results indicate that larger widths of delay distribution increase the regions of amplitude death in the parameter space.
In the case of a uniformly distributed delay kernel, for sufficiently large width of the delay distribution it is possible to achieve amplitude death for an arbitrary value of the average time delay, provided that the coupling strength has a value in the appropriate range. For a gamma distribution of delay, amplitude death is also possible for an arbitrary value of the average time delay, provided that it exceeds a certain value as determined by the coupling phase and the power law of the distribution. The coupling phase has a destabilizing effect and reduces the regions of amplitude death.
\end{abstract}

\section{Introduction}

Coupled oscillator systems are used to model a wide range of physical and biological phenomena, where the dynamics of the underlying complex system is dominated by the interactions of its elements over multiple temporal and spatial scales. The main advantage of using such a modelling approach stems from  the fact that it allows one to obtain vital  information about the behaviour of the complex systems by studying the laws and rules  that govern different dynamical regimes, including (de)synchronization, cooperation, stability etc. For instance, some brain pathologies and cognitive deficiencies, such as Parkinson's disease, Alzheimer's disease, epilepsy, autism are known to be associated with the synchronisation of oscillating neural populations \cite{US06,PHT05,SG05}. Phase synchronisation arises when two or more coupled elements start to oscillate with the same frequency, and if the coupling strength is quite weak, the amplitudes stay unchanged. However, when the coupling strength becomes stronger, the amplitudes start to interact,  which may lead to complete synchronization or amplitude death \cite{CFHBS07,FIE09}.  The coupling between elements is often not instantaneous accounting, for example, for propagation delays or processing times and these time delays have been shown to play a crucial role in sychronisation as well as in (de)stabilisation of general coupled oscillator systems \cite{PRK01,JUS10,DVDF10,CHO09,FYDS10,SHFD10}, coupled semiconductor lasers \cite{HFEMM01,FLU09,HIC11}, neural \cite{DHPS09,SCH08} and engineering systems \cite{KBGHW06,KH10}. 

In coupled oscillator systems with delayed couplings one of the intriguing effects of time delays is amplitude death  \cite{RSJ98,RSJ99}, or death by delay \cite{strogatz98}, when oscillations are suppressed, and the system is driven to a stable equilibrium. The amplitude death phenomenon has been studied both theoretically and experimentally \cite{HFRPO00,TFE00}, and it has been shown that in the presence of time delays, amplitude death can occur even if the frequencies of the individual oscillators are the same. On the contrary, in the non-delayed case, the amplitude death in coupled oscillator systems can only be observed if the frequencies of the individual oscillators are sufficiently different \cite{AEK90,MS90}.

The majority of research on the effects of time delays upon the stability of coupled oscillators has been focussed on systems with one or several constant time delays. This is a valid, though somewhat limiting, assumption for some systems where the time delay is fixed and does not change with time. However, many real-life applications involve time delays which are non-constant \cite{GJU08,GJU10}, or more crucially, where the exact value of the time delay is not explicitly known. This assumption can be overcome by considering distributed-delay systems, where the time delay is given by an integral with a memory kernel in the form of a prescribed delay distribution function. In population dynamics models, distributed time delays have been used to represent maturation time, which varies between individuals \cite{GS03,FT10}; in some engineering systems, the use of distributed time delays is better suited due to the fact that only an approximate value of time delay is known \cite{KK11,MVAN05}; in neural systems, the time delay is different for different paths of the feedback signals and inclusion of a distribution over time delays has been shown to increase the stability region of the system as opposed to the same systems with constant time delays \cite{TSE03}; a similar effect has been shown to occur in  predator-prey and ecological food webs \cite{ETF05}, in epidemiological models distributed time delays have been used to model the effects of waning immunity times after disease or vaccination, which differs between individuals \cite{BK10}. The influence of distributed delays on the overall stability of the system under consideration has been discussed by a number of authors; for example, in \cite{Atay03}, the effects of the time delays are studied in relation to the coupled oscillators, and in traffic dynamics \cite{SAN08}, and in \cite{CJ09} the authors have shown how to approximate the stability boundary around a steady state for a general distribution function.

In this paper we consider a generic system of coupled oscillators with distributed-delay coupling, where the local dynamics is represented by the normal form close to a supercritical Hopf bifurcation. Consider a system of two coupled Stuart-Landau oscillators
\begin{eqnarray}\label{SL}
\dot{z}_1(t)&=&(1+i\omega_1)z_{1}(t)-|z_1(t)|^2z_1(t)\nonumber \\
\nonumber \\
&+&Ke^{i\theta}\left[\int_{0}^{\infty}g(t')z_{2}(t-t')dt'-z_1(t)\right],\nonumber\\ \\
\dot{z}_2(t)&=&(1+i\omega_2)z_{2}(t)-|z_2(t)|^2z_2(t)\nonumber\\ \nonumber\\
&+&Ke^{i\theta}\left[\int_{0}^{\infty}g(t')z_{1}(t-t')dt'-z_2(t)\right],\nonumber
\end{eqnarray}
where $z_{1,2}\in\mathbb{C}$, $\omega_{1,2}$ are the oscillator frequencies, $K\in\mathbb{R}_{+}$ and $\theta\in\mathbb{R}$ are the strength and the phase of coupling, respectively, and $g(\cdot)$ is a distributed-delay kernel, satisfying
\[
g(u)\geq 0,\hspace{0.5cm} \int_{0}^{\infty}g(u)du=1.
\]
When $g(u)=\delta(u)$, one recovers an instantaneous coupling $(z_2-z_1)$; when $g(u)=\delta(u-\tau)$, the coupling takes the form of a discrete time delay $[z_2(t-\tau)-z_1(t)]$. We will concentrate on the case of identical oscillators having the same frequency $\omega_1=\omega_2=\omega_0$. Without coupling, the local dynamics exhibits an unstable steady state $z=0$ and a stable limit cycle with $|z(t)|=1$.
It is known that time delay in the coupling can introduce amplitude death, which means destruction of a periodic orbit and stabilization of the unstable steady state. 

The system (\ref{SL}) has been analysed by Atay in the case of zero coupling phase for a uniformly distributed delay kernel \cite{Atay03}. He has shown that distributed delays increase stability of the steady state and lead to merging of death islands in the parameter space. In this paper, we extend this work in three directions. First of all, we also take into consideration a coupling phase, which is important not only theoretically, but also in experimental realizations of the coupling, as has already been demonstrated in laser experiments \cite{SHWSH,FAYSL}. Second, to get a better understanding of the system behavior inside the stability regions, we will numerically compute  eigenvalues. Finally, we will also consider the case of a practically important gamma distributed delay kernel to illustrate that it is not only the mean delay and the width of the distribution, but also the actual shape of the distribution that affects amplitude death in systems with distributed-delay coupling.

The outline of this paper is as follows. In Sec. \ref{UD} we study amplitude death in the system (\ref{SL}) with a uniformly distributed delay kernel. This includes finding analytically boundaries of stability of the trivial steady state, as well as numerical computation of the eigenvalues of the corresponding characteristic equations. Section \ref{GaD} is devoted to the analysis of amplitude death for the case of a gamma distributed delay kernel. We illustrate how regions of amplitude death are affected by the coupling parameters and characteristics of the delay distribution. The paper concludes with a summary of our findings, together with an outlook on their implications.

\section{Uniformly distributed delay}\label{UD}

To study the possibility of amplitude death in the system (\ref{SL}), we linearize this system near the trivial steady state $z_{1,2}=0$. The corresponding characteristic equation is given by
\begin{equation}\label{ch_eq}
\left(1+i\omega_0 -Ke^{i\theta}-\lambda\right)^{2}-K^{2}e^{2i\theta}\left[\{\mathcal{L}g\}(\lambda)\right]^{2}=0,
\end{equation}
where $\lambda$ is an eigenvalue of the Jacobian, and
\begin{equation}
\{\mathcal{L}g\}(s)=\int_{0}^{\infty}e^{-su}g(u)du,
\end{equation}
is the Laplace transform of the function $g(u)$. To make further analytical progress, it is instructive to specify a particular choice of the delay kernel.
As a first example, we consider a uniformly distributed kernel
\begin{equation}\label{UKer}
g(u)=\left\{
\begin{array}{l}
\displaystyle{\frac{1}{2\rho} \hspace{1cm}\mbox{for }\tau-\rho\leq u\leq \tau+\rho,}\\\\
0\hspace{1cm}\mbox{elsewhere.}
\end{array}
\right.
\end{equation}
This distribution has the mean time delay 
\[
\tau_{m}\equiv<\tau>=\int_{0}^{\infty}ug(u)du=\tau,
\]
and the variance
\begin{equation}\label{VD}
\displaystyle{\sigma^2=\int_{0}^{\infty}(u-\tau_m)^{2}g(u)du=\frac{\rho^{2}}{3}.}
\end{equation}
In the case of a uniformly distributed kernel Eq.~(\ref{UKer}), it is quite easy to compute the Laplace transform of the distribution $g(u)$ as:
\[
\{\mathcal{L}g\}(\lambda)=\frac{1}{2\rho\lambda}e^{-\lambda\tau}\left(e^{\lambda\rho}-e^{-\lambda\rho}\right)=e^{-\lambda\tau}
\frac{\sinh(\lambda\rho)}{\lambda\rho},
\]
and this also transforms the characteristic equation (\ref{ch_eq}) as
\begin{equation}\label{ch_eq_2}
1+i\omega_0 -Ke^{i\theta}-\lambda=\pm Ke^{i\theta}e^{-\lambda\tau}
\frac{\sinh(\lambda\rho)}{\lambda\rho}.
\end{equation}
Since the roots of the characteristic equation (\ref{ch_eq_2}) are complex-valued, stability of the trivial steady state can only change if some of these eigenvalues cross the imaginary axis.
To this end, we can look for characteristic roots in the form $\lambda=i\omega$. Substituting this into the characteristic equation (\ref{ch_eq_2}) and separating real and imaginary parts gives the following system of equations for $(K,\tau)$:
\begin{equation}\label{Ktau}
\begin{array}{l}
\displaystyle{K^2\left[1-\delta(\rho,\omega)\right]-2K[\cos\theta+(\omega_0-\omega)\sin\theta]}\\\\
\hspace{2.5cm}+(\omega_0-\omega)^2+1=0,\\\\
\displaystyle{\tan(\theta-\omega\tau)=\frac{\omega_0-\omega-K\sin\theta}{1-K\cos\theta},}
\end{array}
\end{equation}
where
\[
\delta(\rho,\omega)=\left[\frac{\sin(\omega\rho)}{\omega\rho}\right]^2,
\]
We begin the analysis of the effect of the coupling phase $\theta$ on stability by first considering the case $\theta=0$. In this case, the system (\ref{Ktau}) simplifies to
\begin{equation}\label{T0Ktau}
\begin{array}{l}
\displaystyle{K^2\left[1-\delta(\rho,\omega)\right]-2K+(\omega_0-\omega)^2+1=0,}\\\\
\displaystyle{\tan(\omega\tau)=\frac{\omega-\omega_0}{1-K}.}
\end{array}
\end{equation}

To illustrate the effects of varying the coupling stren gth $K$ and the time delay $\tau$ on the (in)stability of the trivial steady state, we now compute the stability boundaries (\ref{T0Ktau}) as parametrized by the Hopf frequency $\omega$. Besides the stability boundaries themselves, which enclose the amplitude death regions, we also compute the maximum real part of the eigenvalues using the traceDDE package in Matlab.
In order to compute these eigenvalues, we introduce real variables $z_{1r,i}$ and $z_{2r,i}$, where $z_1=z_{1r}+iz_{1i}$ and $z_2=z_{2r}+iz_{2i}$, and rewrite the linearized system (SL) with the distributed kernel (\ref{UKer}) as
\begin{equation}\label{Trace}
\displaystyle{\dot{\bf z}(t)=L_0 {\bf z}(t)+\frac{K}{2\rho}\int_{-(\tau+\rho)}^{-(\tau-\rho)}M{\bf z}(t+s)ds,}
\end{equation}
where
\[
\begin{array}{c}
{\bf z}=(z_{1r},z_{1i},z_{2r},z_{2i})^{T},\hspace{0.3cm}
L_0=\left(
\begin{array}{ll}
N&{\bf 0}_{2}\\
{\bf 0}_{2}&N
\end{array}
\right),\\\\
M=\left(
\begin{array}{ll}
{\bf 0}_{2}&R\\
R&{\bf 0}_{2}
\end{array}
\right),\hspace{0.5cm}
R=\left(
\begin{array}{ll}
\cos\theta&\sin\theta\\
-\sin\theta&\cos\theta
\end{array}
\right),\\\\
N=\left(
\begin{array}{ll}
1-K\cos\theta&K\sin\theta-\omega_0\\
\omega_0-K\sin\theta&1-K\cos\theta
\end{array}
\right),
\end{array}
\]
and ${\bf 0}_{2}$ denotes a $2\times 2$ zero matrix. When $\rho=0$, the last term in the system (\ref{Trace}) turns into $KM{\bf z}(t-\tau)$, which describes the system with a single discrete time delay $\tau$. System (\ref{Trace}) is in the form in which it is amenable to the algorithms described in Breda {\it et al.} \cite{BMS06} and implemented in traceDDE.

\begin{figure*}
\hspace{-0.5cm}
\includegraphics[width=18cm]{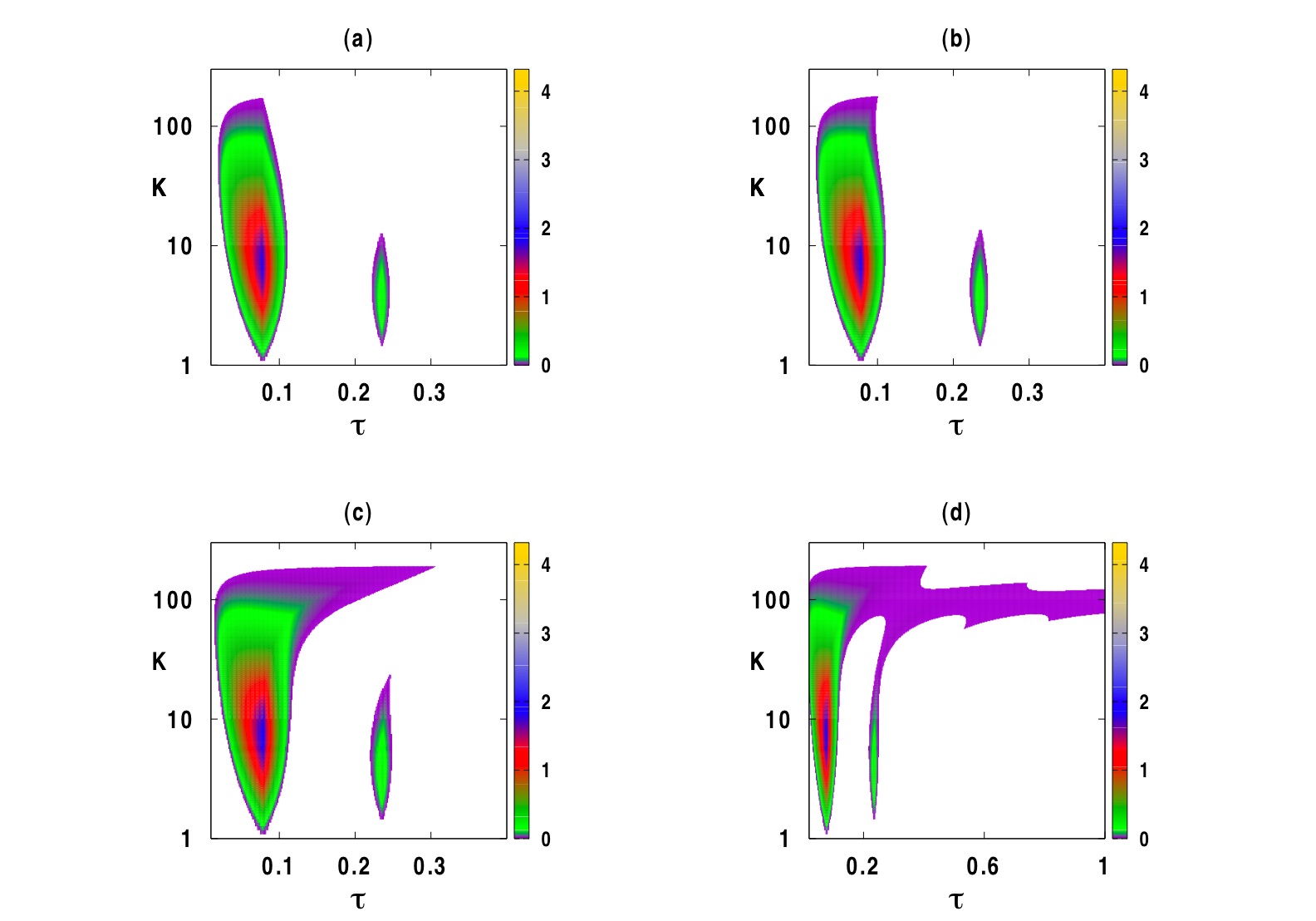}
\caption{Areas of amplitude death in the plane of the coupling strength $K$ and the mean time delay $\tau$ for $\theta=0$, $\omega_0=20$. Colour code denotes $[-\max\{{\rm Re}(\lambda)\}]$ for $\max\{{\rm Re}(\lambda)\}\le 0$. (a) $\rho=0$. (b) $\rho=0.005$. (c) $\rho=0.015$. (d) $\rho=0.018$.}\label{ADfig}
\end{figure*}

Figure~\ref{ADfig} shows the boundaries of the amplitude death together with the magnitude of the real part of the leading eigenvalue for different widths of the delay distribution $\rho$.
When $\rho=0$ (single discrete time delay $\tau$), there are two distinct islands in the ($K,\tau$) parameter space, in which the trivial steady state is stable. It is noteworthy that the number of such stability islands increases with increasing fundamental frequency $\omega_0$ (e.g., there are three islands for $\omega_0=30$).
\begin{figure*}
\hspace{-0.5cm}
\includegraphics[width=18cm]{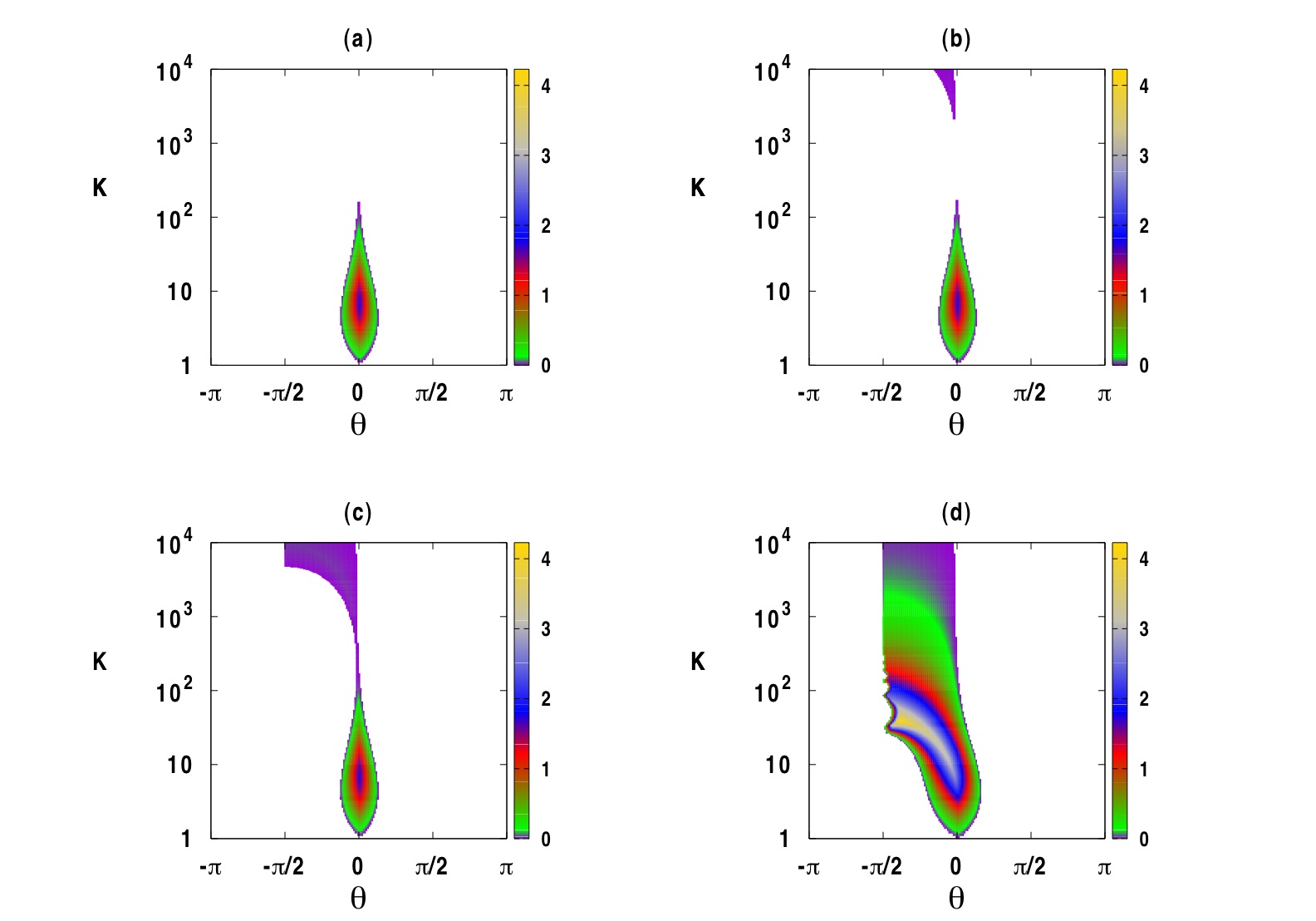}
\caption{Areas of amplitude death depending on the coupling strength $K$ and the phase $\theta$ for $\tau=0.08$ and $\omega_0=20$. Colour code denotes $[-\max\{{\rm Re}(\lambda)\}]$. (a) $\rho=0$. (b) $\rho=0.002$. (c) $\rho=0.004$. (d) $\rho=0.026$.}\label{KTheta}
\end{figure*}
This behaviour agrees with that of an equivalent single system with time-delayed feedback and delay $2\tau$ \cite{HOE05}, where similar islands of stability of the trivial steady state form in the ($K,\tau$) plane around $2\tau= \frac{2n+1}{2} T_0 \equiv \frac{2n+1}{2} 2\pi/\omega_0$ ($n=0,1,2,...$),
and the number and size of those islands increases with increasing ratio $\omega_0/\lambda_0$ \cite{footnote1}. As the width of the delay distribution $\rho$ increases, the stability islands grow (see Fig.~\ref{ADfig} (b) and (c)) until they merge into a single continuous region in the parameter space, as shown in Fig.~\ref{ADfig} (d).

Next, we consider the effects of the coupling phase on stability of the trivial steady state in the system (\ref{SL}) with a uniformly distributed delay kernel (\ref{UKer}). The boundaries of amplitude death in this case as parametrized by the Hopf frequency $\omega$ are given in Eq.~(\ref{Ktau}). Figure~\ref{KTheta} shows how the areas of amplitude death depend on the coupling strength and coupling phase for various delay distribution widths $\rho$. In the case of a single delay ($\rho=0$), the only area of amplitude death is symmetric around $\theta=0$ and also has the largest range of $K$ values at $\theta=0$, for which amplitude death is achieved, as illustrated in Fig.~\ref{KTheta}(a). Provided that the coupling phase is large enough by the absolute value, oscillations in the system (\ref{SL}) are maintained, and amplitude death cannot be achieved for any values of the coupling strength $K$. As the width of the delay distribution $\rho$ increases, this leads to an increase in the size of the amplitude death region in ($K,\theta$) space, as well as to an asymmetry with regard to the coupling phase: for sufficiently large $\rho$, amplitude death can occur for an arbitrarily large $K$ if $\theta$ is negative, but only for a very limited range of $K$ values if $\theta$ is positive. At the same time, if the coupling phase exceeds $\pi/2$ by the absolute value, amplitude death does not occur, irrespectively of the values of $K$.

\begin{figure*}
\hspace{-0.5cm}
\includegraphics[width=17cm]{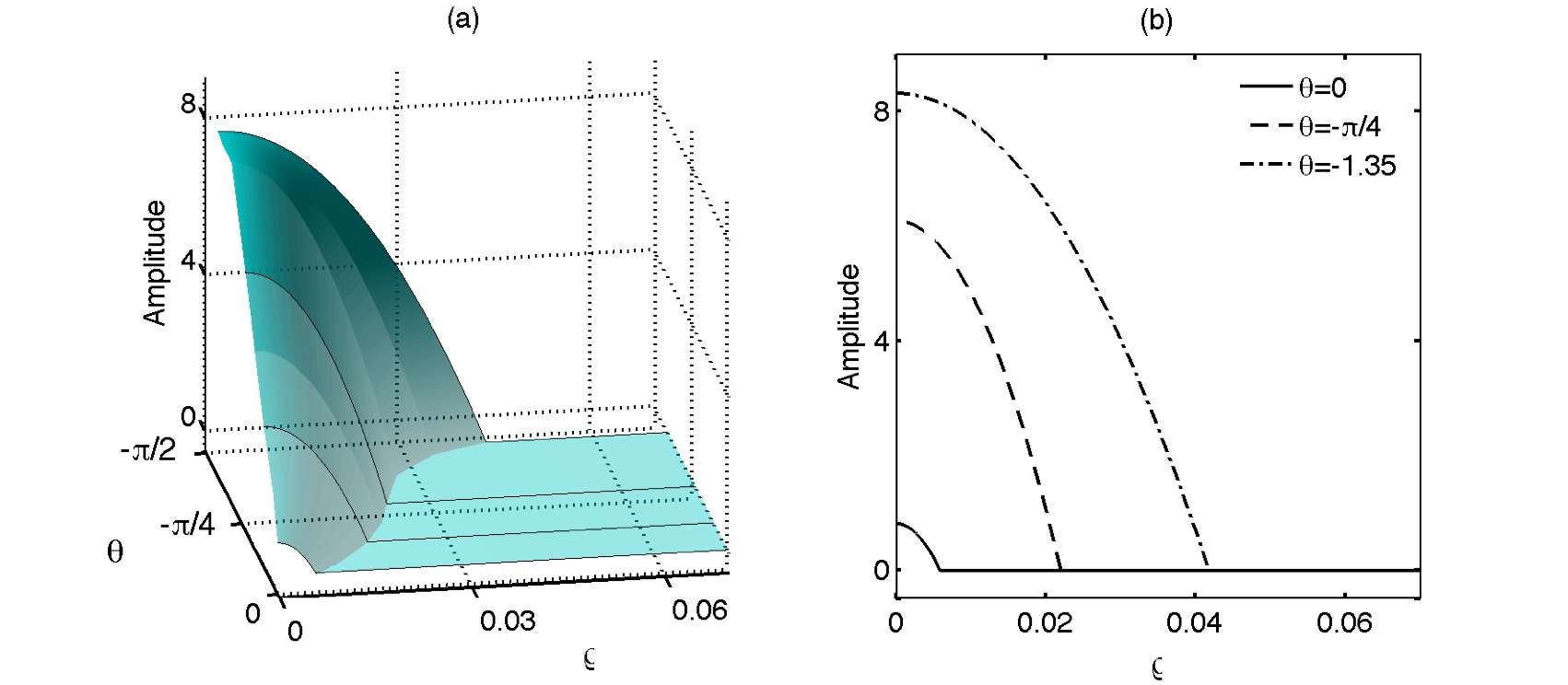}
\caption{(a) Amplitude of oscillations depending on the coupling phase $\theta$ and the width of distribution $\rho$. (b) Sections for various values of $\theta$. Parameter values are $K=120$, $\tau=0.08$ and $\omega_0=20$.}\label{Amp_plot}
\end{figure*}

Whilst distributed-delay coupling may fail to suppress oscillations in a certain part of the parameter space, it will still have an effect on the amplitude of oscillations. In Fig.~\ref{Amp_plot} we illustrate how the amplitude of oscillations varies depending on the coupling phase $\theta$ and the delay distribution width $\rho$ for a fixed value of the coupling strength $K$. As the magnitude of the coupling phase increases, the range of possible delay distribution widths, for which oscillations are observed, is growing, and the amplitude reaches its maximum at $\rho=0$ corresponding to the case of discrete time delay.

\section{Gamma distributed delay}\label{GaD}

In many realistic situations, the distribution of time delays is better represented by a gamma distribution, which can be written as
\begin{equation}
g(u)=\frac{u^{p-1}\alpha^{p}e^{-\alpha u}}{\Gamma(p)},
\end{equation}
with $\alpha,p\geq 0$, and $\Gamma(p)$ being an Euler gamma function defined by $\Gamma(0)=1$ and $\Gamma(p+1)=p\Gamma(p)$. For integer powers $p$, this can be equivalently written as
\begin{equation}\label{GD}
g(u)=\frac{u^{p-1}\alpha^{p}e^{-\alpha u}}{(p-1)!}.
\end{equation}
For $p=1$ this is simply an exponential distribution (also called a {\it weak delay kernel}) with the maximum contribution to the coupling coming from the present values of variables $z_{1}$ and $z_{2}$. For $p>1$ (known as {\it strong delay kernel} in the case $p=2$), the biggest influence on the coupling at any moment of time $t$ is from the values of $z_{1,2}$ at $t-(p-1)/\alpha$.

The delay distribution (\ref{GD}) has the mean time delay
\begin{equation}
\displaystyle{\tau_{m}=\int_{0}^{\infty}ug(u)du=\frac{p}{\alpha},}
\end{equation}
and the variance
\[
\displaystyle{\sigma^2=\int_{0}^{\infty}(u-\tau_m)^{2}g(u)du=\frac{p}{\alpha^2}}.
\]
When studying stability of the trivial steady state of the system (\ref{SL}) with the delay distribution kernel (\ref{GD}), one could use the same strategy as the one described in the previous section. The only complication with such an analysis would stem once again from the Laplace transform of the distribution kernel, which in this case has the form
\[
\{\mathcal{L}g\}(\lambda)=\frac{\alpha^{p}}{(\lambda+\alpha)^{p}}.
\]
A convenient way to circumvent this is to use the {\it linear chain trick} \cite{MD78}, which allows one to replace an equation with a gamma distributed delay kernel by an equivalent system of $(p+1)$ ordinary differential equations. To illustrate this, we consider a particular case of system (\ref{SL}) with a weak delay kernel given by (\ref{GD}) with $p=1$, which is equivalent to a low-pass filter \cite{HOE05}:
\begin{equation}\label{WK}
g_{w}(u)=\alpha e^{-\alpha u}.
\end{equation}
Introducing new variables
\[
\begin{array}{l}
\displaystyle{Y_{1}(t)=\int_{0}^{\infty}\alpha e^{-\alpha s}z_{1}(t-s)ds,}\\\\
\displaystyle{Y_{2}(t)=\int_{0}^{\infty}\alpha e^{-\alpha s}z_{2}(t-s)ds,}
\end{array}
\]
allows us to rewrite the system (\ref{SL}) as follows
\begin{eqnarray}\label{ED_sys}
\dot{z}_1(t)&=&(1+i\omega_1)z_{1}(t)-|z_1(t)^2|z_1(t)\nonumber \\
\nonumber \\
&+&Ke^{i\theta}\left[Y_2(t)-z_1(t)\right],\nonumber\\
\nonumber \\
\dot{z}_2(t)&=&(1+i\omega_2)z_{2}(t)-|z_2(t)^2|z_2(t)\nonumber \\
\nonumber \\
&+&Ke^{i\theta}\left[Y_1(t)-z_2(t)\right],\nonumber\\\\
\dot{Y}_{1}(t)&=&\alpha z_{1}(t)-\alpha Y_{1}(t),\nonumber\\
\nonumber\\
\dot{Y}_{2}(t)&=&\alpha z_{2}(t)-\alpha Y_{2}(t),\nonumber
\end{eqnarray}
where the distribution parameter $\alpha$ is related to the mean time delay as $\alpha=1/\tau_{m}$. The trivial equilibrium $z_1=z_2=0$ of the original system (\ref{SL}) corresponds to a steady state $z_1=z_2=Y_1=Y_2=0$ of the modified system (\ref{ED_sys}).
The characteristic equation for the linearization of system (\ref{ED_sys}) near this trivial steady state reduces to
\begin{eqnarray}\label{CEQ}
&&[\lambda^2+\lambda\left(Ke^{i\theta}-1+\alpha-i\omega_0\right)-\alpha(1+i\omega_0)]\times\nonumber\\
\nonumber\\
&&[\lambda^2+\lambda\left(Ke^{i\theta}-1+\alpha-i\omega_0\right)\\
\nonumber\\
&&-\alpha(1+i\omega_0-2Ke^{i\theta})]=0.\nonumber
\end{eqnarray}
\begin{figure*}
\hspace{-0.5cm}
\includegraphics[width=9cm]{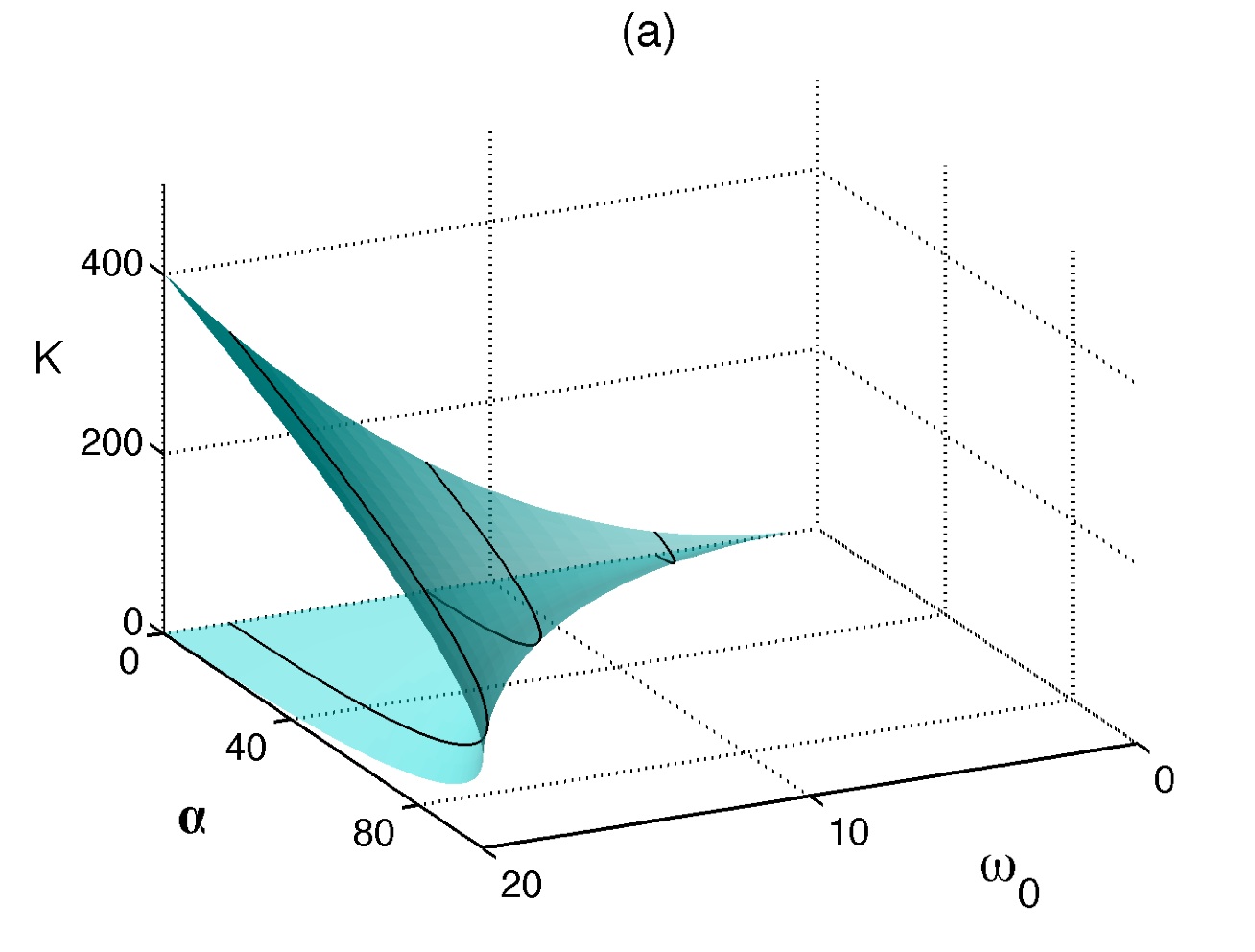}\hspace{-0.8cm}
\includegraphics[width=10.5cm]{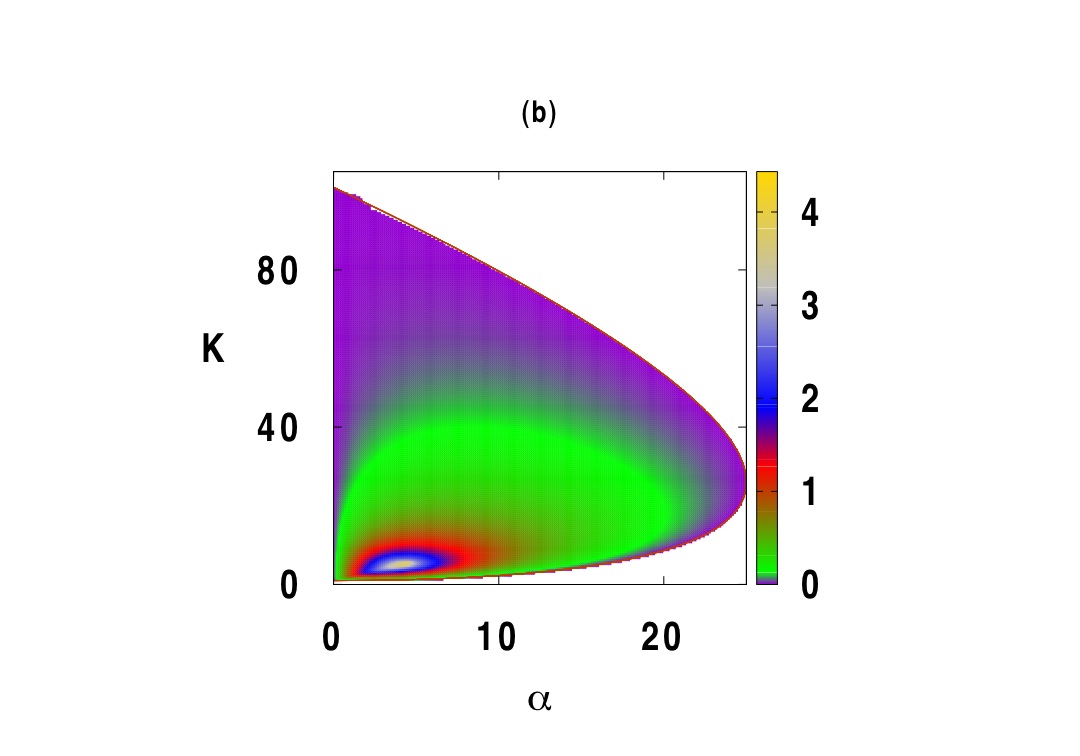}
\caption{(a) Stability boundary for the system (\ref{SL}) with a weak delay distribution kernel (\ref{WK}) for $\theta=0$ ($p=1$).The trivial steady state is unstable outside the boundary surface and stable inside the boundary surface. (b) Stability boundary for $\omega_0=10$. Colour code denotes $[-\max\{{\rm Re}(\lambda)\}]$.}\label{EB0}
\end{figure*}
Let us consider the first factor in Eq.~(\ref{CEQ})
\begin{equation}\label{lam1}
\lambda^2+\lambda\left(Ke^{i\theta}-1+\alpha-i\omega_0\right)-\alpha(1+i\omega_0)=0.
\end{equation}
Since $\lambda=0$ is not a solution of this equation, the only way how stability of the trivial steady state can change is when $\lambda$ crosses the imaginary axis. To find the values of system parameters when this can happen, we look for solutions of equation (\ref{lam1}) in the form $\lambda=i\omega$. Substituting this into the above equation and separating real and imaginary parts gives
\[
\begin{array}{l}
\omega K\sin\theta=\omega\omega_0-\alpha-\omega^2,\\\\
\omega K\cos\theta=\alpha\omega_0+\omega(1-\alpha).
\end{array}
\]
Solving this system gives the coupling strength  $K$ and the inverse mean time delay $\alpha$ as functions of the coupling phase $\theta$ and the Hopf frequency $\omega$:
\begin{equation}\label{EBoundT}
\begin{array}{l}
\displaystyle{K=\frac{1+(\omega-\omega_{0})^{2}}{\cos\theta+\sin\theta(\omega_0-\omega)},}\\\\
\displaystyle{\alpha=-\frac{\omega[\sin\theta+\cos\theta(\omega-\omega_{0})}{\cos\theta+\sin\theta(\omega_0-\omega)}}.
\end{array}
\end{equation}
When the coupling phase $\theta$ is equal to zero, these expressions simplify to
\begin{equation}\label{EBound0}
K=1+(\omega-\omega_0)^2,\hspace{0.5cm}\alpha=\omega(\omega_0-\omega).
\end{equation}
This implies that the permissible range of Hopf frequencies is $0\leq\omega\leq\omega_0$, the minimal value of the mean time delay for which the amplitude death can occur is $\tau_{\rm min}=1/\alpha_{\rm max}=4/\omega_{0}^{2}$, and the minimal coupling strength required for amplitude death is $K=1$. Figure~\ref{EB0} illustrates how the stability boundary (\ref{EBound0}) depends on the intrinsic frequency $\omega_0$, and it also shows how the leading eigenvalues vary inside the stable parameter region. As the intrinsic frequency $\omega_0$ increases, the region in the ($K,\alpha$) space for which amplitude death occurs grows, thus indicating that it is possible to stabilise a trivial steady state for even smaller values of the mean time delay.

\begin{figure*}
\hspace{-1cm}
\includegraphics[width=17cm]{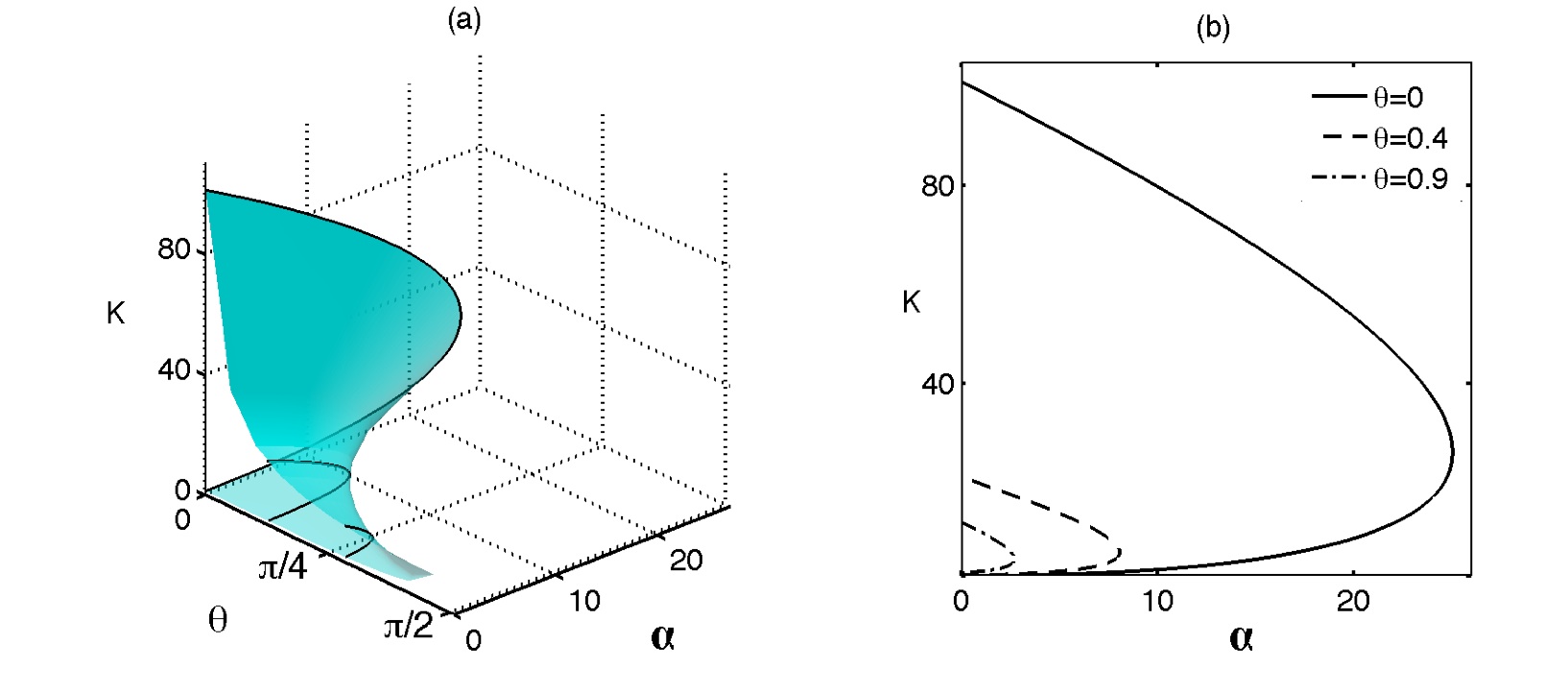}
\caption{(a) Stability boundary for the system (\ref{SL}) with a weak delay distribution kernel (\ref{WK}) with $\omega_0=10$ ($p=1$). The trivial steady state is unstable outside the boundary and stable inside the boundary. (b) Sections for various values of $\theta$.}\label{EBT}
\end{figure*}

\begin{figure*}\hspace{-0.2cm}
\includegraphics[width=8cm]{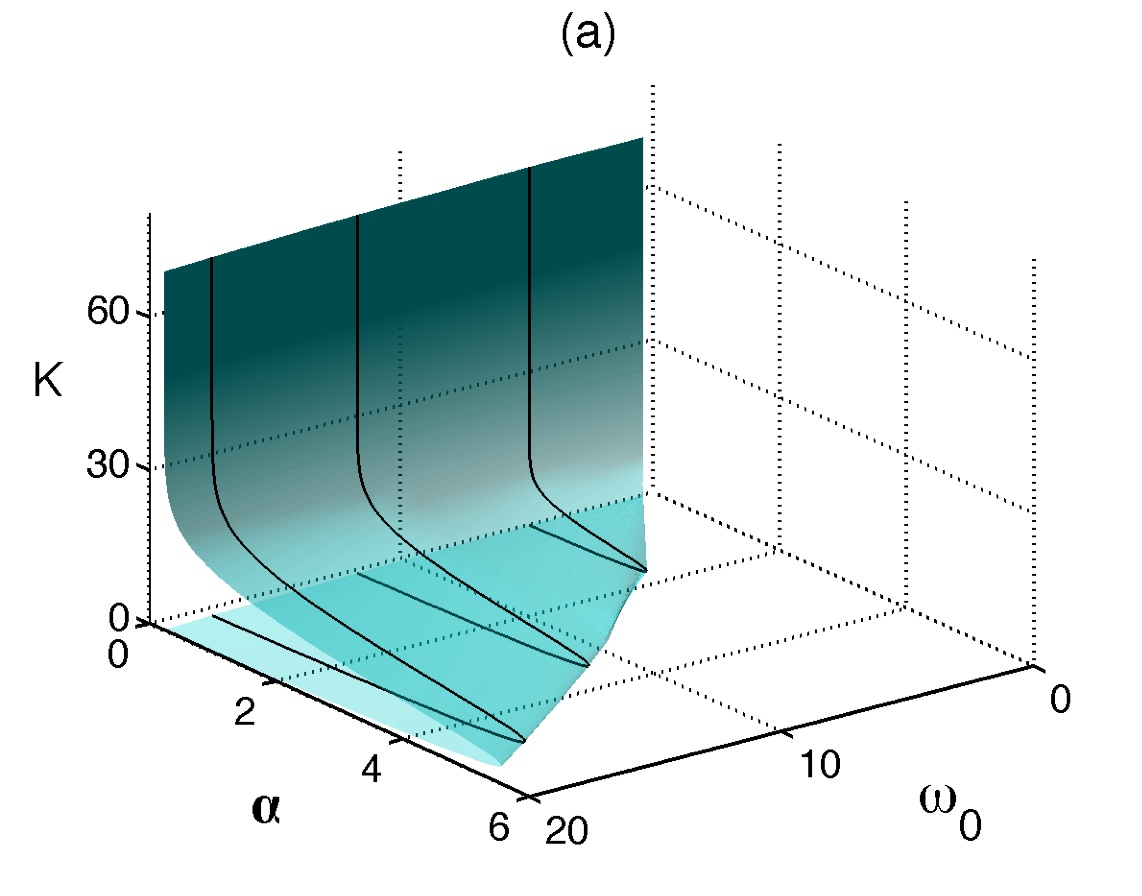}\hspace{-0.4cm}
\includegraphics[width=10.5cm]{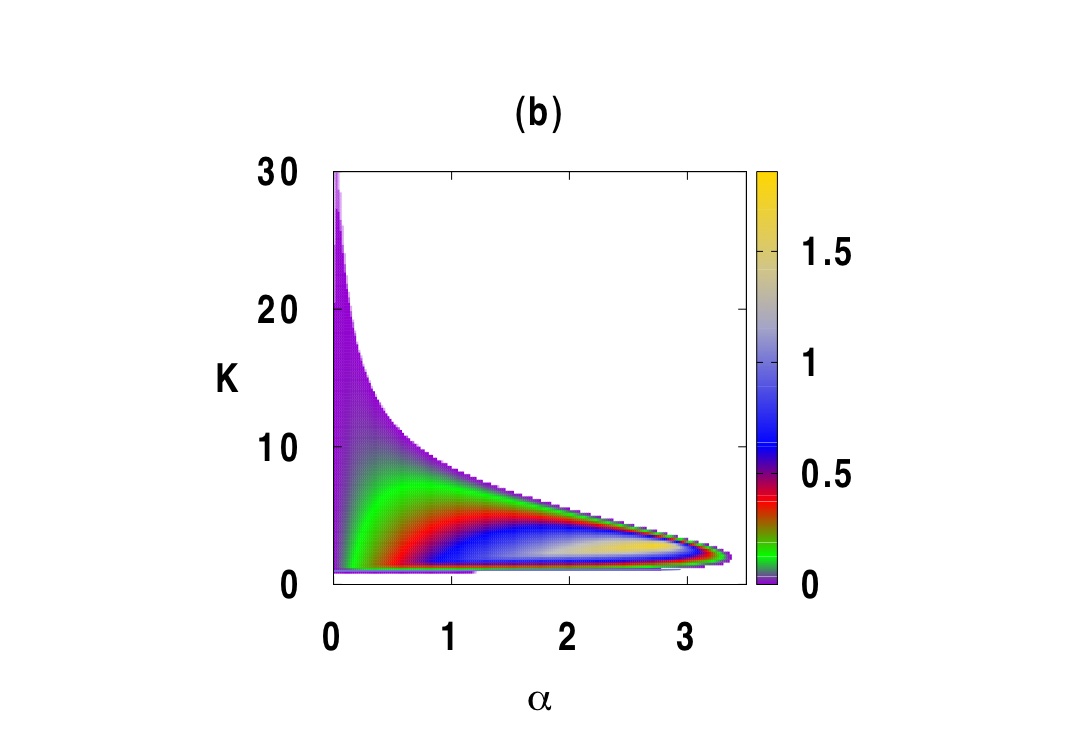}
\caption{(a) Stability boundary for the system (\ref{SL}) with a strong delay distribution kernel (\ref{SK}) for $\theta=0$ ($p=2$). The trivial steady state is unstable outside the boundary surface and stable inside the boundary surface. (b) Stability boundary for $\omega_0=10$. Colour code denotes $[-\max\{{\rm Re}(\lambda)\}]$.}\label{SKB}
\end{figure*}

For $\theta$ different from zero, the requirement $K\geq 0$, $\alpha\geq 0$ in (\ref{EBoundT}) translates into a restriction on admissible coupling phases $0\leq\theta\leq\arctan(\omega_0)$. Figure~\ref{EBT} shows how for a fixed value of $\omega_0$ the stability area reduces as $\theta$ grows, eventually collapsing at $\theta=\arctan(\omega_0)$. The range of admissible Hopf frequencies also reduces and is given for each $\theta$ by $0\leq\omega\leq\omega_0-\tan\theta$. The second factor in (\ref{CEQ}) provides another stability boundary in the parameter space, but the values of $K$ and $\alpha$ satisfying this equation with $\lambda=i\omega$ lie outside the feasible range of $K\geq 0$, $\alpha\geq 0$. Hence, it suffices to consider the stability boundary given by (\ref{EBoundT}).

\begin{figure*}
\hspace{-1cm}
\includegraphics[width=17cm]{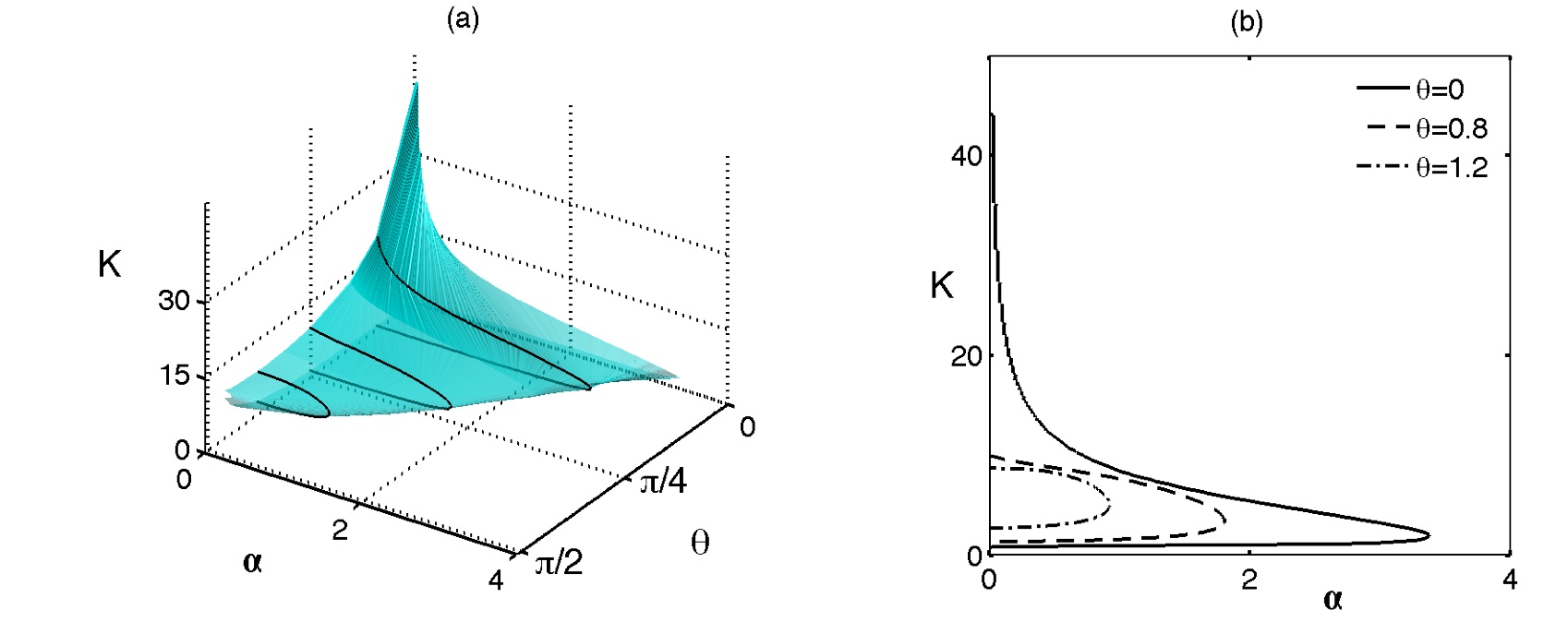}
\caption{(a) Stability boundary for the system (\ref{SL}) with a strong delay distribution kernel (\ref{SK}) with $\omega_0=10$ ($p=2$). The trivial steady state is unstable outside the boundary and stable inside the boundary. (b) Sections for various values of $\theta$.
}\label{SKtheta}
\end{figure*}

Next, we consider the case of the strong delay kernel ($p=2$)
\begin{equation}\label{SK}
g_{w}(u)=\alpha^{2}u e^{-\alpha u}.
\end{equation}
Following the same strategy as in the case of the weak delay kernel (\ref{ED_sys}), we introduce new variables
\[
\begin{array}{l}
Y_{11}(t)=\int_{0}^{\infty}\alpha e^{-\alpha s}z_{1}(t-s)ds,\\\\
Y_{12}(t)=\int_{0}^{\infty}\alpha^{2}s e^{-\alpha s}z_{1}(t-s)ds,\\\\
Y_{21}(t)=\int_{0}^{\infty}\alpha e^{-\alpha s}z_{2}(t-s)ds,\\\\
Y_{22}(t)=\int_{0}^{\infty}\alpha^{2}s e^{-\alpha s}z_{2}(t-s)ds,
\end{array}
\]
and then rewrite the system (\ref{SL}) in the form
\begin{eqnarray}\label{SK_sys}
\dot{z}_1(t)&=&(1+i\omega_1)z_{1}(t)-|z_1(t)^2|z_1(t)\nonumber\\
\nonumber \\
&+&Ke^{i\theta}\left[Y_{12}(t)-z_1(t)\right],\nonumber \\
\nonumber\\
\dot{z}_2(t)&=&(1+i\omega_2)z_{2}(t)-|z_2(t)^2|z_2(t)\nonumber\\
\nonumber \\
&+&Ke^{i\theta}\left[Y_{22}(t)-z_2(t)\right],\nonumber \\\\
\dot{Y}_{11}(t)&=&\alpha z_{1}(t)-\alpha Y_{11}(t),\nonumber\\
\nonumber \\
\dot{Y}_{12}(t)&=&\alpha^2 z_{1}(t)+\alpha Y_{11}(t)-\alpha Y_{12}(t),\nonumber\\
\nonumber\\
\dot{Y}_{21}(t)&=&\alpha z_{2}(t)-\alpha Y_{21}(t),\nonumber\\
\nonumber\\
\dot{Y}_{22}(t)&=&\alpha^2 z_{2}(t)+\alpha Y_{21}(t)-\alpha Y_{22}(t),\nonumber
\end{eqnarray}
where the mean time delay is given by $\tau_{m}=2/\alpha$. Linearizing system (\ref{SK_sys}) near the steady state $z_1=z_2=Y_{11}=Y_{12}=Y_{21}=Y_{22}=0$ yields a characteristic equation, which is more involved than in the case of the weak delay kernel. Solving this characteristic equation provides the boundary of amplitude death depending on system parameters. Figure~\ref{SKB} illustrates how this boundary changes with the coupling strength $K$, fundamental frequency $\omega_0$, and the inverse time delay $\alpha/2$ in the case $\theta=0$. Similar to the case of the weak delay kernel, as the inverse time delay $\sim \alpha$ approaches zero, the larger value of the critical coupling strength of $K$ 
tends to some finite value. At the same time, the lower value of the critical coupling strength $K$ does vary depending on the coupling phase unlike the case of the weak kernel. Also, in the case of a strong delay kernel, the amplitude death is achieved for a much smaller range of coupling strengths $K$ and only for much larger mean time delays $\tau_{m}=2/\alpha$. However, the effect of the coupling phase on the region of amplitude death is similar to that for a weak delay kernel, i.e., as the coupling phase increases, this reduces the area in the $(K,\alpha)$ parameter plane where amplitude death is observed, as shown in Fig.~\ref{SKtheta}.

\section{Discussion}\label{Disc}

In this paper, we have studied the effects of distribu- ted-delay coupling on the stability 
of the trivial steady state in a system of coupled oscillators. Using generic Stuart-Landau oscillators, we have identified parameter regimes of amplitude death, i.e., stabilised steady states, in terms of intrinsic oscillator frequency and characteristics of the coupling. In order to better understand the dynamics inside stable regimes, we have numerically computed eigenvalues of the corresponding characteristic equation. We have considered two particular types of delay distribution: a uniform distribution around some mean time delay and a class of exponential or more general gamma distributions. For both of these distributions it has been possible to find stability in terms of coupling strength, coupling phase, and the mean time delay.
These results suggest that the coupling phase plays an important role in determining the ranges of admissible Hopf frequencies and values of the coupling strength, for which stabilisation of the trivial steady state is possible.

For the uniformly distributed delay kernel, as the width of the distribution increases, the region of amplitude death in the parameter space of the coupling strength and the average time delay increases. As the coupling phase $\theta$ moves away from zero, the range of coupling strength providing amplitude death gets smaller until some critical value of the width of distribution, beyond which an asymmetry in $\theta$ is introduced and this range is significantly larger for negative values of $\theta$ than it is for $\theta\geq 0$. Furthermore, as the magnitude of the coupling phase grows, the maximum amplitude of oscillations in the coupled system also grows, and such oscillations can be observed for a larger range of delay distribution widths.

In the case of gamma distributed delay, the region of amplitude death does not consist of isolated stability islands but is always a continuous region in the $(K,\tau_{m})$ plane. Similar to the uniform delay distribution, as the coupling phase increases from zero, the range of coupling strengths for which stabilisation of the steady state can be achieved reduces, while the minimum average time delay required for the stabilization increases. However, unlike the uniform distribution,  within the stable region the amplitude death can occur for an arbitrarily large value of the average time delay, provided it is above some minimum value and the coupling strength is within the appropriate range.

So far, we have considered the effects of distribu- ted-delay coupling on the dynamics of identical coupled oscillators only. The next step would be to extend this analysis to the case when the oscillators have differing intrinsic frequencies in order to understand the dynamics of the phase difference, as well as to analyse the stability of in-phase and 
anti-phase oscillations.

\section*{Acknowledgements}

This work was partially supported by DFG in the framework of SFB 910: {\em Control of self-organizing nonlinear
systems: Theoretical methods and concepts of application}.


\begin{thebibliography}{50}

\bibitem{US06} P.~J.~Uhlhaas and W.~Singer, {\it Neuron} {\bf 52}, 155 (2006).

\bibitem{PHT05} O.~V.~Popovych, C.~Hauptmann, and P.~A.~Tass, {\it Phys. Rev. Lett.} {\bf 94}, 164102 (2005).

\bibitem{SG05} A.~Schnitzler and J.~Gross,  {\it Nat. Rev. Neuroscience} {\bf 6}, 285 (2005).

\bibitem{CFHBS07} C.~U.~Choe, V.~Flunkert, P.~H\"ovel, H.~Benner, and E.~Sch\"oll, {\it Phys. Rev. E} {\bf 75}, 0426206 (2007).

\bibitem{FIE09} B.~Fiedler, V.~Flunkert, P.~H{\"o}vel, and E.~Sch{\"o}ll, {\it Phil.
  Trans.~R. Soc.~A} {\bf 368}, 319 (2010).

\bibitem{PRK01} A.~Pikovsky, M.~Rosenblum, and J.~Kurths, {\it Synchronization: a universal concept in nonlinear sciences}. (CUP, Cambridge, 2001).

\bibitem{JUS10}
W.~Just, A.~Pelster, M.~Schanz, and E.~Sch{\"o}ll (eds.), {\it Phil. Trans.~R. Soc.~A} {\bf 368}, 303 (2010).

\bibitem{DVDF10} O.~D'Huys, R.~Vicente, J.~Danckaert, and I.~Fischer, {\it Chaos} {\bf 20}, 043127 (2010).

\bibitem{CHO09}
C.~U. Choe, T.~Dahms, P.~H{\"o}vel, and E.~Sch{\"o}ll, {\it Phys. Rev.~E} {\bf 81}, 025205(R) (2010).

\bibitem{FYDS10}  V.~Flunkert, S. Yanchuk, T. Dahms, and E. Sch\"oll, {\it Phys. Rev. Lett.} {\bf 105}, 254101 (2010).

\bibitem{SHFD10} E.~Sch\"oll, P.~H\"ovel, V.~Flunkert, and M.~A. Dahlem, in {\it Complex Time-Delay Systems}, edited by F.~M. Atay (Springer, Berlin, 2010), p. 85.

\bibitem{HFEMM01} T.~Heil, I.~Fischer, W.~Els\"asser,  J.~Mulet, and C.~R.~Mirasso, {\it Phys. Rev. Lett} {\bf 86}, 795 (2001).

\bibitem{FLU09}
V.~Flunkert, O.~{D'Huys}, J.~Danckaert, I.~Fischer, and E.~Sch{\"o}ll, {\it Phys. Rev.~E} {\bf 79}, 065201 (R)
  (2009).

\bibitem{HIC11} K.~Hicke, O.~{D'Huys}, V.~Flunkert, E.~Sch{\"o}ll, J.~Danckaert, and
  I.~Fischer, {\it Phys. Rev.~E} {\bf 83}, 056211 (2011).

\bibitem{DHPS09} M.~A. Dahlem, G.~Hiller, A.~Panchuk, and E.~Sch\"oll, {\it Int. J. Bifur. Chaos} {\bf 19},  745 (2009).

\bibitem{SCH08}
E.~Sch{\"o}ll, G.~Hiller, P.~H{\"o}vel, and M.~A. Dahlem, {\it Phil. Trans.~R. Soc.~A} {\bf 367}, 1079 (2009).

\bibitem{KBGHW06} Y.~N. Kyrychko, K.~B. Blyuss, A.~Gonzalez-Buelga, S.~J. Hogan \& D.~J.~Wagg, {\it Proc. R. Soc. A} {\bf 462}, 1271 (2006).

\bibitem{KH10} Y.~N. Kyrychko and S.~J. Hogan, {\it J. Vibr. Control} {\bf 16}, 943 (2010).

\bibitem {RSJ98} D.~V. Ramana Reddy, A.~Sen, and G.~L.~Johnston, {\it Phys. Rev. Lett.} {\bf 80}, 5109 (1998).

\bibitem{RSJ99} D.~V. Ramana Reddy, A.~Sen, and G.~L.~Johnston, {\it Physica D} {\bf 129}, 15 (1999).

\bibitem{strogatz98} S.~H.~Strogatz, {\it Nature} {\bf 394}, 316 (1998).

\bibitem{HFRPO00} R.~Herrero, M. Figueras, J. Rius, F. Pi, and G. Orriols,  {\it Phys. Rev. Lett.} {\bf 84}, 5312 (2000).

\bibitem{TFE00} A. Takamatsu, T. Fujii, and I. Endo,  {\it Phys. Rev. Lett.} {\bf 85}, 2026 (2000).

\bibitem{AHL04} A.~Ahlborn, U.~Parlitz, {\it Phys. Rev. Lett.} {\bf 93}, 264101 (2004).

\bibitem{AHL05} A.~Ahlborn, U.~Parlitz, {\it Phys. Rev. E} {\bf 72}, 016206 (2005).

\bibitem{AEK90} D.~G. Aronson, G.~B. Ermentrout, and N.~Kopell, {\it Physica D} {\bf 41}, 403 (1990).

\bibitem{MS90} R.~E. Mirollo and S.~H. Strogatz, {\it J. Stat. Phys.} {\bf 60}, 245 (1990).

\bibitem{GJU08} A.~Gjurchinovski and V.~Urumov, {\it Europhys. Lett.} {\bf 84}, 40013 (2008).

\bibitem{GJU10} A.~Gjurchinovski and V.~Urumov, {\it Phys. Rev. E} {\bf 81}, 016209 (2010).

\bibitem{GS03} S.~A.~Gourley \& J.~W.-H. So,  {\it Proc. R. Soc. Edinburgh} {\bf 133}, 527 (2003).

\bibitem{FT10} T.~Faria and S.~Trofimchuk, {\it Nonlinearity} {\bf 23}, 2457 (2010).

\bibitem{KK11} G.~Kiss and B.~Krauskopf, {\it Dyn. Syst.} {\bf 26}, 85 (2011).

\bibitem{MVAN05} W.~Michiels, V.~Van Assche, and S.-I. Niculescu, {\it IEEE Trans. Automat. Contr.} {\bf 50}, 493 (2005).

\bibitem{TSE03} A.~Thiel, H.~Schwegler, and C.~W.~Eurich, {\it Complexity} {\bf 8}, 102 (2003).

\bibitem{ETF05} C.~W. Eurich, A.~Thiel, and L.~Fahse, {\it Phys. Rev. Lett.} {\bf 94}, 158104 (2005).

\bibitem{BK10} K.~B. Blyuss and Y.~N. Kyrychko, {\it Bull. Math. Biol.} {\bf 72}, 490 (2010). 

\bibitem{Atay03} F.~Atay, {\it Phys. Rev. Lett.} {\bf 91},  094101 (2003).

\bibitem{SAN08} R.~Sipahi, F.~M.~Atay, and S.-I.~Niculescu, {\it SIAM J. Appl. Math.} {\bf 68}, 738 (2008).

\bibitem{CJ09} S.~A.~Campbell and R.~Jessop, {\it Math. Model. Nat. Phenom.} {\bf 4}, 1 (2009).

\bibitem{SHWSH} S.~Schikora, P.~H\"ovel, H.-J.~W\"unsche, E.~Sch\"oll, and F. Henneberger, {\it Phys. Rev. Lett.} {\bf 97}, 213902 (2006).

\bibitem{FAYSL} A.~P.~A.~Fischer, O.~K.~Andersen, M.~Yousefi, S.~Stolte, and D.~Lenstra,
{\it IEEE J. Quantum Electron.} {\bf 36}, 375 (2000).

\bibitem{BMS06} D.~Breda, S.~Maset, and R.~Vermiglioa, {\it Appl. Numer. Math.} {\bf 56}, 318 (2006).

\bibitem{footnote1}
Note that we have set the bifurcation parameter $\lambda_0=1$ in Eq. (1) and throughout the present paper.

\bibitem{MD78} N.~MacDonald, {\it Time lags in biological systems}. (Springer, New York, 1978).

\bibitem{HOE05}
P.~H{\"o}vel and E.~Sch{\"o}ll, {\it Phys.~Rev.~E} {\bf 72}, 046203 (2005).

\end{thebibliography}
\end{document}